\documentclass[12pt,a4paper]{iopart}
\usepackage{graphicx,iopams,color,cite}

\begin{document}

\title[Spatial averaging and gradient sensing]{Role of spatial averaging in multicellular gradient sensing}

\author{Tyler Smith}
\address{Department of Physics, Emory University, Atlanta, GA 30322, USA}

\author{Sean Fancher}
\address{Department of Physics and Astronomy, Purdue University, West Lafayette, IN 47907, USA}

\author{Andre Levchenko}
\address{Department of Biomedical Engineering and Yale Systems Biology Institute, Yale University, New Haven, CT 06520, USA}

\author{Ilya Nemenman}
\address{Departments of Physics and Biology, Emory University, Atlanta, GA 30322, USA}
\ead{ilya.nemenman@emory.edu}

\author{Andrew Mugler}
\address{Department of Physics and Astronomy, Purdue University, West Lafayette, IN 47907, USA}
\ead{amugler@purdue.edu}

\begin{abstract} Gradient sensing underlies important biological
  processes including morphogenesis, polarization, and cell migration.
  The precision of gradient sensing increases with the length of a
  detector (a cell or group of cells) in the gradient direction, since
  a longer detector spans a larger range of concentration values.
  Intuition from analyses of concentration sensing suggests that
  precision should also increase with detector length in the direction
  transverse to the gradient, since then spatial averaging should
  reduce the noise. However, here we show that, unlike for
  concentration sensing, the precision of gradient sensing decreases
  with transverse length for the simplest gradient sensing model,
  local excitation--global inhibition (LEGI). The reason is that gradient sensing
  ultimately relies on a subtraction of measured concentration values.
  While spatial averaging indeed reduces the noise in these
  measurements, which increases precision, it also reduces the
  covariance between the measurements, which results in the net
  decrease in precision. We demonstrate how a recently
  introduced gradient sensing mechanism, regional excitation--global
  inhibition (REGI), overcomes this effect and recovers the benefit of
  transverse averaging. Using a REGI-based model, we compute the
  optimal two- and three-dimensional detector shapes, and argue that
  they are consistent with the shapes of naturally occurring
  gradient-sensing cell populations. \end{abstract}

\pacs{00.00, 00.00}

\vspace{2pc}
\noindent{\it Keywords}: gradient sensing, cell-cell communication, spatial averaging


\section{Introduction}

Determining the strength and direction of a chemical concentration
gradient is an essential task for a diverse array of biological
processes. Gradient sensing underlies the polarization of single
cells, the orientation and migration of cells and cell collectives,
and the changes in tissue morphology that occur during embryogenesis
and the subsequent development of an organism
\cite{Rosoff2004,Onsum:2006gg,Song2006,sternlicht2006hormonal,Bianco:2007hi,Friedl:2009kz,mani2013collective,Dona:2013cd,Pocha:2014ep,huebner2014cellular,MaletEngra:2015jc,ellison2015cell}.
Experiments have shown that cells are remarkably precise gradient
sensors \cite{Rosoff2004,ellison2015cell}, and a large amount of
effort has gone into understanding the mechanisms of, and the limits
to, biological gradient sensing
\cite{Berg1977,Goodhill1999,Levchenko2002,Onsum:2006gg,Endres2008,Endres2009,Hu2010,Jilkine:2011fy,mugler2015limits}.

At its core, gradient sensing requires the comparison of concentration
measurements between the ``front'' and the ``back'' of a detector.
Front and back here are defined with respect to the gradient
direction, and the detector here is a single cell or a group of cells.
If the front and back are more separated, then the concentration
measurements are more different from each other, which improves the
determination of the gradient. This implies that detectors that are
longer in the gradient direction have a higher gradient sensing
precision \cite{Goodhill1999,Endres2008,Endres2009,Hu2010}. This
argument neglects the fact that information must be communicated
between different parts of a detector, especially if the detector is
multicellular. Recently we derived the limits to the precision of
gradient sensing including communication, and we found that for a
one-dimensional (1-D) detector, the precision indeed increases with
detector length, but then saturates due to the fact that communication
introduces its own noise \cite{ellison2015cell,mugler2015limits}.
Nonetheless, the precision of gradient sensing increases or saturates
with the length of a 1-D detector aligned with the gradient; it does
not decrease.

Yet biological detectors are not 1-D in general.  Two-dimensional
(2-D) detectors include the quasi-cylindrical arrangement of cell
nuclei during the early stages of Drosophila development
\cite{Gregor:2007du} and the planar arrangement of epithelial cell
layers \cite{mani2013collective}. Three-dimensional (3-D) detectors
include single cells and the multicellular tips of growing epithelial
ducts \cite{ewald2008collective}, as well as border cells collective
guidance in Drosophila \cite{Bianco:2007hi}. This raises the question
of what effect the dimensions transverse to the gradient direction
have on the precision of gradient sensing.

Intuition about this question can be drawn from the similar task of
sensing the value of a concentration (as opposed to sensing its
difference between two points in space, i.~e., the gradient). If the
concentration profile is uniform in space, then the precision of
concentration sensing benefits from increasing the detector length in
any direction. The reason is that communication with other parts of
the detector, or {\it spatial averaging}, does not change the mean of
a particular measurement within the detector, but it does reduce the
noise \cite{Berg1977,Endres2008,Endres2009}. Even if the concentration
profile is graded, but the goal is still concentration (rather than
gradient) sensing, as in stripe formation in early Drosophila
development, the precision still benefits from spatial averaging
\cite{erdmann2009role}.\footnote{The distinction between gradient
  sensing, and concentration sensing with a graded profile, is a
  subtle but important one, and is further discussed in Results
  section 1 and the Discussion.} The benefit is especially clear in a
direction transverse to the gradient direction: once again, spatial
averaging in this direction does not change the mean of a particular
measurement, but it does reduce the noise. These considerations, drawn
from the problem of concentration sensing, suggest that the precision
of gradient sensing should also increase with the length of a detector
in a direction transverse to the gradient.

Here we investigate theoretically and computationally the precision of
gradient sensing for 2-D and 3-D detectors. We start with one of the
simplest models of gradient sensing, the local excitation--global
inhibition (LEGI) model \cite{Levchenko2002,Jilkine:2011fy}. This is an
accepted basic model when gradient sensing is adaptive (that is,
 background concentration largely does not effect the gradient
sensing). Surprisingly, in contrast to the case of concentration
sensing, we find that the precision of gradient sensing decreases with
the length of the detector in a direction transverse to the gradient
direction. The reason is that gradient sensing fundamentally relies on
a subtraction of concentration measurements, e.g.\ between the front
and back of the detector. While spatial averaging reduces the
intrinsic noise in these measurements, which increases precision, it
also reduces the covariance between the measurements, which decreases
precision. We demonstrate that the latter effect dominates, such that
the net result is a decrease in precision with transverse detector
size. Then we show that this decrease can actually be overcome by a
gradient-sensing strategy that we recently introduced, termed regional
excitation--global inhibition (REGI) \cite{mugler2015limits}. We
demonstrate that REGI retains a high covariance between measurements
and restores the benefit of transverse averaging. Using a REGI-based
model, we compute the optimal 2-D and 3-D detector shapes, which arise
from an interplay of the effects of transverse averaging on both the
signal and the noise of gradient detection. We argue that these shapes
are consistent with the shapes of the multicellular tips of epithelial
ducts, suggesting that this and other similarly shaped
gradient-sensing systems benefit from spatial averaging in all
dimensions.

\section{Background}

\begin{figure}
\centering
\includegraphics[width=.9\textwidth]{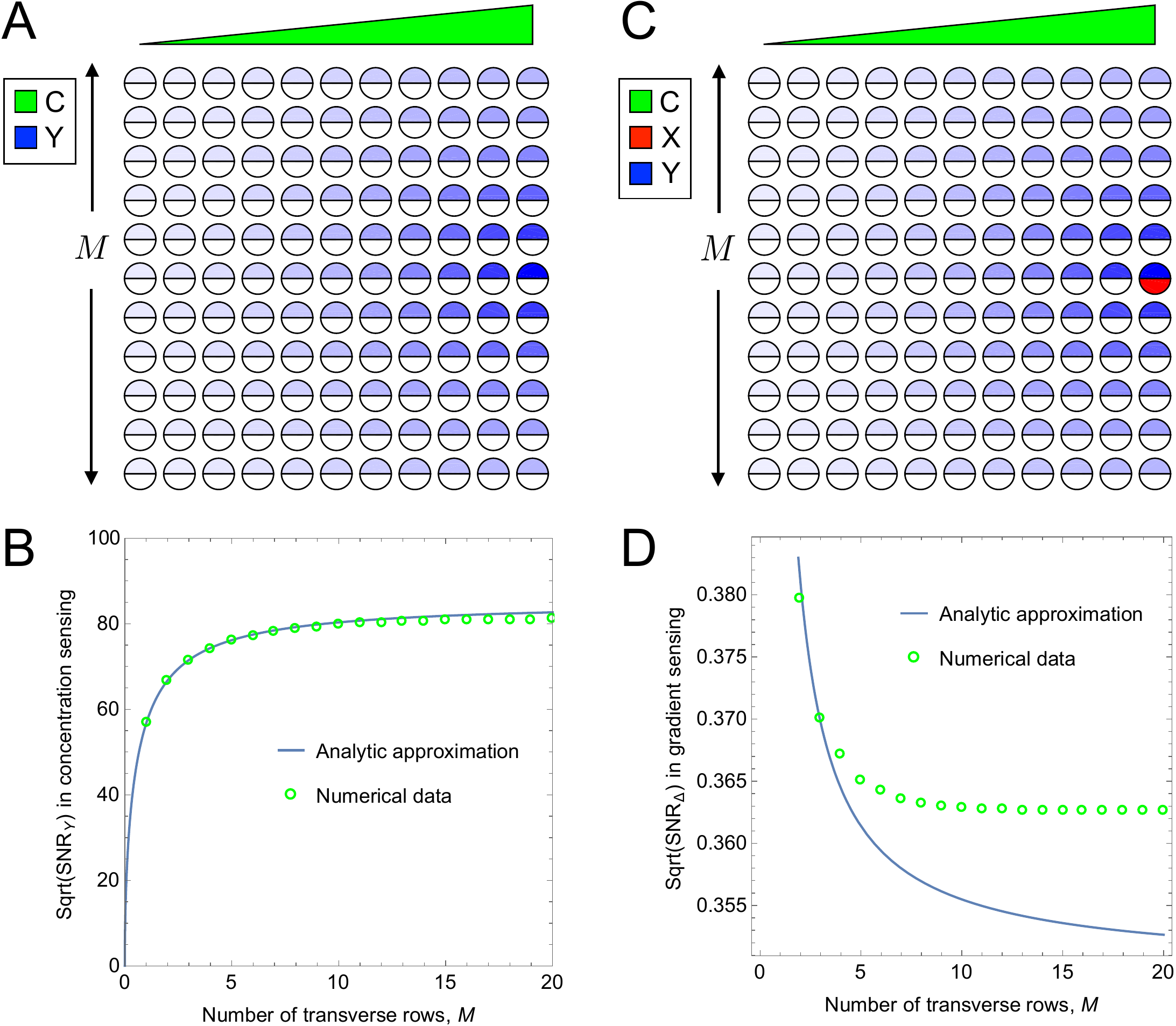}
\caption{\label{transverse} Spatial averaging transverse to a gradient
  improves concentration sensing, but worsens gradient sensing. (A) A
  2-D array of cells is exposed to a concentration profile C that
  varies linearly in the horizontal direction (green wedge). In each
  cell, Y molecules are produced in proportion to the local C value. Y
  molecules are also exchanged between neighboring cells, providing
  the spatial averaging. Thus Y is the readout for the average
  concentration in the vicinity of a particular cell. Blue indicates
  the mean number of Y molecules $\bar{y}$ in each cell that have
  originated from the rightmost, middle cell. (B) The signal-to-noise
  ratio (SNR) for $y$ increases with the number $M$ of rows of cells
  added transverse to the gradient direction. (C) As in A, but with an
  additional internal species X. The molecules are also produced in
  proportion to the local C value, but they are not exchanged between
  cells. Red indicates the mean number of X molecules $\bar{x}$ in
  each cell that have originated from the rightmost, middle cell. The
  difference $\Delta = x-y$ provides the readout for the gradient
  (LEGI). (D) In contrast to B, the SNR for $\Delta$ decreases with
  the number of transverse rows $M$. In B and D, the numerical results
  are compared with the theoretical approximations (see Eqs.\
  \ref{eq:conc_a} and \ref{eq:grad_a}, respectively) and agree at
  small $M$ as expected. Parameters are similar to the experiments in
  \cite{ellison2015cell}: $\bar{c}_N = 1.25$ nM, $g = 0.5$ nM/mm,
  $a = 10$ $\mu$m, $n_y = 4$, $N = 50$ cells per row, and $G = 10$.}
\end{figure}

As in previous work \cite{ellison2015cell, mugler2015limits}, we
consider the local excitation--global inhibition (LEGI) model of
multicellular gradient sensing, which is a minimal, adaptive,
spatially extended model of gradient sensing. We consider a signal
concentration profile $c$ that varies linearly in a particular
direction in 3-D space, with concentration gradient $g$
(Fig.\ \ref{transverse}A, C). In the $n$th cell, both a local
molecular species X and a global molecular species Y are produced at a
rate $\beta$ and degraded at a rate $\mu$. The production rate is also
proportional to the number of signal molecules in the cell's vicinity
$c_na^3$, where $a$ is the cell diameter. Whereas the local species X
is confined to each cell, the global species Y is exchanged between
neighboring cells at a rate $\gamma_y$ (Fig.\ \ref{transverse}C).
Conceptually, X measures the local concentration of signal molecules,
while Y represents their spatially-averaged concentration. As in
\cite{ellison2015cell, mugler2015limits} we consider the linear
response regime, in which the dynamics of the local and global species
satisfy the stochastic equations
\begin{eqnarray}
\label{eq:dxdt}
\frac{dx_n}{dt} &=& \beta (c_na^3) - \mu x_n + \eta_n, \\
\frac{dy_n}{dt} &=& \beta (c_na^3) - \mu y_n + \gamma_y \sum_{n'\in{\cal N}(n)} (y_{n'}-y_n) + \xi_n \nonumber \\
\label{eq:dydt}
&=& \beta (c_na^3) - \mu \sum_{n'} M^y_{nn'} y_{n'} + \xi_n.
\end{eqnarray}
Here
$M^y_{nn'} \equiv (1+|{\cal N}_n|\gamma_y/\mu)\delta_{nn'} -
(\gamma_y/\mu)\delta_{n'\in{\cal N}_n}$
is the connectivity matrix for the global species that accounts for
degradation and molecule exchange. ${\cal N}_n$ and $|{\cal N}_n|$
denote the indices and the number of nearest neighbors of cell $n$,
respectively. The intrinsic noise terms $\eta_n$ and $\xi_n$
correspond to the Poissonian production, degradation, and exchange
reactions \cite{ellison2015cell}.

In the LEGI paradigm, X excites a downstream species while Y inhibits
it. If the cell is at the higher edge of the gradient, then the local
concentration (X) is higher than the spatial average (Y), and the
excitation exceeds the inhibition. While such comparison of the
excitation and the inhibition can be done by many different molecular
mechanisms \cite{Jilkine:2011fy}, we consider here the limit of
shallow gradients, where the comparison is equivalent to subtracting Y
from X \cite{ellison2015cell}. This difference,
$\Delta_n = x_n - y_n$, is the readout of the model. If
$\Delta_n$ is positive, the $n$th cell is further up the gradient than
average; if $\Delta_n$ is negative, the $n$th cell is further down the
gradient than average. In this work, we always focus on the readout
$\Delta_N$ of the cell highest up the gradient, which we denote as the
$N$th cell.

We assume that the cells do not average concentrations of the signal C
and the messenger molecules X and Y over time (though generalizations
with averaging are certainly possible \cite{mugler2015limits}). Then
the precision of gradient sensing is given by the square root of the
instantaneous signal-to-noise ratio (SNR) for the readout,
SNR$_\Delta = (\bar{\Delta}_N/\delta\Delta_N)^2$, where the mean and
variance are given by \cite{ellison2015cell}
\begin{eqnarray}
\bar{\Delta}_N &=& \bar{x}_N - \bar{y}_N, \\
\label{eq:xbar}
\bar{x}_N &=& Ga^3\bar{c}_N, \label{xgain}\\
\bar{y}_N &=& Ga^3 \sum_n K^y_n \bar{c}_{N-n},\label{ygain}
\end{eqnarray}
and
\begin{eqnarray}
(\delta\Delta_N)^2 &=& (\delta x_N)^2 + (\delta y_N)^2 - 2{\rm cov}(x_N, y_N), \\
\label{eq:varx}
(\delta x_N)^2 &=& \bar{x}_N + G^2a^3 \bar{c}_N, \\
\label{eq:vary}
(\delta y_N)^2 &=& \bar{y}_N + G^2 a^3 \sum_n (K^y_n)^2 \bar{c}_{N-n}, \\
\label{eq:cov}
{\rm cov}(x_N, y_N) &=& G^2a^3 K^y_0 \bar{c}_N,
\end{eqnarray}
respectively. Here $K^y_n \equiv (M^y)^{-1}_{N,N-n}$ is the
communication kernel, and $G \equiv \beta/\mu$ is the gain. The first
terms in Eqs.\ \ref{eq:varx} and \ref{eq:vary} correspond to intrinsic
noise, while the second terms correspond to extrinsic noise and assume
that the diffusion of the signal is slow \cite{ellison2015cell}.
Computing the precision for a given configuration of cells only
requires inverting the connectivity matrix $M^y$.

In a 1-D geometry, and in the limit of many cells
($N\gg 1$) and fast communication ($\gamma_y \gg \mu$), the kernel
reduces to $K^y_n \approx e^{-n/n_y}/n_y$, where
$n_y \equiv \sqrt{\gamma_y/\mu}$ sets the effective length scale of
communication \cite{ellison2015cell}. In this limit, the variance in
the global species and the covariance reduce to \cite{ellison2015cell} 
\begin{eqnarray} \label{eq:vary_a}
  (\delta y_N)^2 &\approx& \bar{y}_N + G^2 a^3 \frac{\bar{c}_{N-n_y/2}}{2n_y}, \\
  \label{eq:cov_a} {\rm cov}(x_N, y_N) &\approx& G^2a^3
                                                 \frac{\bar{c}_N}{n_y}.
\end{eqnarray}

In the recently introduced regional excitation--global inhibition
(REGI) model \cite{mugler2015limits}, the local species X is also
exchanged among cells, but at a lower rate $\gamma_x < \gamma_y$.
Then Eq.\ \ref{eq:dxdt} becomes analogous to Eq.\ \ref{eq:dydt}, and
Eqs.\ \ref{eq:xbar}, \ref{eq:varx}, and \ref{eq:cov} are replaced by
\begin{eqnarray}
\bar{x}_N &=& Ga^3 \sum_n K^x_n \bar{c}_{N-n}, \\
(\delta x_N)^2 &=& \bar{x}_N + G^2 a^3 \sum_n (K^x_n)^2 \bar{c}_{N-n}, \\
{\rm cov}(x_N, y_N) &=& G^2a^3 \sum_n K^x_n K^y_n \bar{c}_{N-n},
\end{eqnarray}
respectively, where $K^x_n \equiv (M^x)^{-1}_{N,N-n}$ is the
communication kernel for the local species, and
$M^x_{nn'} \equiv (1+|{\cal N}_n|\gamma_x/\mu)\delta_{nn'} -
(\gamma_x/\mu)\delta_{n'\in{\cal N}_n}$.
Once more, computing the precision for a given configuration of cells
in the REGI model only requires inverting the connectivity matrices
$M^x$ and $M^y$. While diffusion of X decreases $\bar{x}_N$ at the
$N$th cell, and hence decreases the difference $\bar{\Delta}_N$, it
also averages X over a larger volume, hence decreasing its noise. As
shown in Ref.~\cite{mugler2015limits}, under a broad range of
conditions, the decrease in the noise dominates, and the overall
precision of the REGI model is higher than that of LEGI.

\section{Results}

\subsection{Concentration sensing precision increases with transverse detector size}

Before investigating gradient sensing, we focus on the simpler problem
of concentration sensing. In the local excitation--global inhibition
(LEGI) model, both X and Y provide readouts of the local
concentration, while their difference $\Delta$ provides a readout of
the gradient. The concentration readout provided by Y is spatially
averaged, whereas the concentration readout provided by X is not. Even
if the signal profile is graded, X and Y are concentration readouts if
viewed independently (with different spatial averaging), not gradient
readouts. For example, during Drosophila development, the morphogen
profiles are graded, but individual nuclei in the embryo measure (and
threshold) the local concentration, possibly with some spatial
averaging
\cite{Gregor:2007du,erdmann2009role,Lander:2011bj,Sokolowski:2015tn}.

How does the precision of concentration sensing depend on transverse
detector size? To answer this question, we focus on the spatially
averaged concentration readout Y. We consider a linear signal profile
with gradient $g$ and compute the SNR of Y in the $N$th cell, as we
vary the number $M$ of rows of cells in a direction transverse to the
gradient (Fig.\ \ref{transverse}A). We see in Fig.\ \ref{transverse}B
(green circles) that the precision of concentration sensing increases
with $M$. The reason is that adding rows of cells transverse to the
gradient allows for Y molecules to be exchanged between rows (in
addition to along each row). This does not change the mean $\bar{y}_N$
due to the translational symmetry in the transverse direction.
However, it does reduce the variance, since the global species Y is
now averaged over more cells. The net effect is an increase in the
SNR beyond what is allowed by longitudinal averaging.

We can elucidate the effect of spatial averaging more quantitatively
by appealing to the expression for the variance in Y in a single row
of cells, in the limit of many cells and fast communication (Eq.\
\ref{eq:vary_a}). For a small number of added rows ($M < n_y$, where
$n_y$ is the lengthscale of the spatial averaging), we make the
approximation that the averaging is nearly uniform over all $M$ rows.
In this case, the intrinsic component of the variance is unchanged
(since the mean is unchanged), but the extrinsic component is reduced
by $M$,
\begin{equation}
\label{eq:conc_a}
(\delta y_N)^2 \approx \bar{y}_N + G^2 a^3 \frac{\bar{c}_{N-n_y/2}}{2n_yM}.
\end{equation} 
The SNR calculated using this approximation is compared with the
numerical result in Fig.\ \ref{transverse}B. We see that the
approximation agrees with the numerical data, and that the agreement
is best for small $M < n_y = 4$, as expected.

\subsection{Gradient sensing precision decreases with transverse detector size}

We now turn our attention to gradient sensing. How does the precision
of gradient sensing depend on transverse detector size? To answer this
question for a linear signal profile, we compute the SNR of the
gradient readout $\Delta_N$ as a function of the number $M$ of rows of
cells in a direction transverse to the gradient (Fig.\
\ref{transverse}C). We see in Fig.\ \ref{transverse}D that the
precision of gradient sensing decreases with $M$ (green circles). This
is in contrast to the precision of concentration sensing, which
increases with $M$ (Fig.\ \ref{transverse}B).

To understand why the precision of gradient sensing decreases with
$M$, we once again consider the mean and the variance of the readout.
The mean $\bar{\Delta}_N = \bar{x}_N - \bar{y}_N$ does not change with
$M$ because neither $\bar{x}_N$ nor $\bar{y}_N$ change with $M$.
However, the variance $(\delta\Delta_N)^2 = (\delta x_N)^2 + (\delta
y_N)^2 - 2{\rm cov}(x_N, y_N)$ changes with $M$ due to two effects.
First, the variance in the global species $(\delta y_N)^2$ decreases
with $M$ due to spatial averaging, as discussed in the previous
section. Second, the covariance ${\rm cov}(x_N, y_N)$ also decreases
with $M$ because Y is exchanged with a larger number of cells, whereas
X is not exchanged, so the two covary more weakly. The effects have
opposite signs. To understand which effect dominates, we again appeal to
the expressions for a single row of cells in the limit of many cells
and fast communication (Eqs.\ \ref{eq:vary_a} and \ref{eq:cov_a}). For
small $M < n_y$, if we make the approximation that the averaging is
nearly uniform over all $M$ rows, then both (i) the extrinsic
component of the variance in Y and (ii) the covariance are reduced by
$M$,
\begin{eqnarray}
(\delta \Delta_N)^2 &\approx& (\delta x_N)^2 + \bar{y}_N + G^2 a^3 \frac{\bar{c}_{N-n_y/2}}{2n_yM}
	- 2G^2a^3 \frac{\bar{c}_N}{n_yM} \nonumber \\
\label{eq:grad_a}
&=& (\delta x_N)^2 + \bar{y}_N - G^2 a^3 \frac{[4\bar{c}_N - \bar{c}_{N-n_y/2}]}{2n_yM}.
\end{eqnarray}
Because the $N$th cell is at the highest concentration, we have
$\bar{c}_N > \bar{c}_{N-n_y/2}$, and we see that Eq.\ \ref{eq:grad_a}
is a decreasing function of $M$. Therefore, the decrease of the
covariance dominates over the decrease of the variance in Y, for all
parameter values. Because the mean $\bar{\Delta}_N$ does not change
with $M$, we conclude that the precision of gradient sensing decreases
with transverse detector size. The SNR calculated using this
approximation is compared with the numerical result in Fig.\
\ref{transverse}D. We see that the approximation agrees with the
numerical data for small $M < n_y = 4$, as expected. The disagreement
at large $M$ is more apparent here than in Fig.\ \ref{transverse}B
because the precision of gradient sensing is low compared to the
precision of concentration sensing, i.~e.\ the gradient is shallow
compared to the background concentration.

\subsection{REGI mechanism recovers the benefit of transverse averaging}

In the previous section we saw that the precision of gradient sensing
using the LEGI model (local messenger X is not exchanged among the
cells) decreases with the size of a detector in a direction transverse
to the gradient, due to the fact that the covariance between the
subtracted variables decreases with the transverse size. For the REGI
model, exchange of the X molecules has an additional effect beyond
increasing the sensing precision for 1-D line of cells
\cite{mugler2015limits}: it increases the covariance of X and Y,
compared to the LEGI mechanism. Indeed, now both X and Y are amplified,
downstream signals from some of the same external ligand molecules.
Since the decrease of gradient sensing precision with transverse
detector size is due to the loss of covariance (Fig.\
\ref{transverse}D), this raises the question of whether the REGI
strategy can overcome this effect and allow gradient sensing precision
to benefit from transverse averaging.

\begin{figure}
\centering
\includegraphics[width=\textwidth]{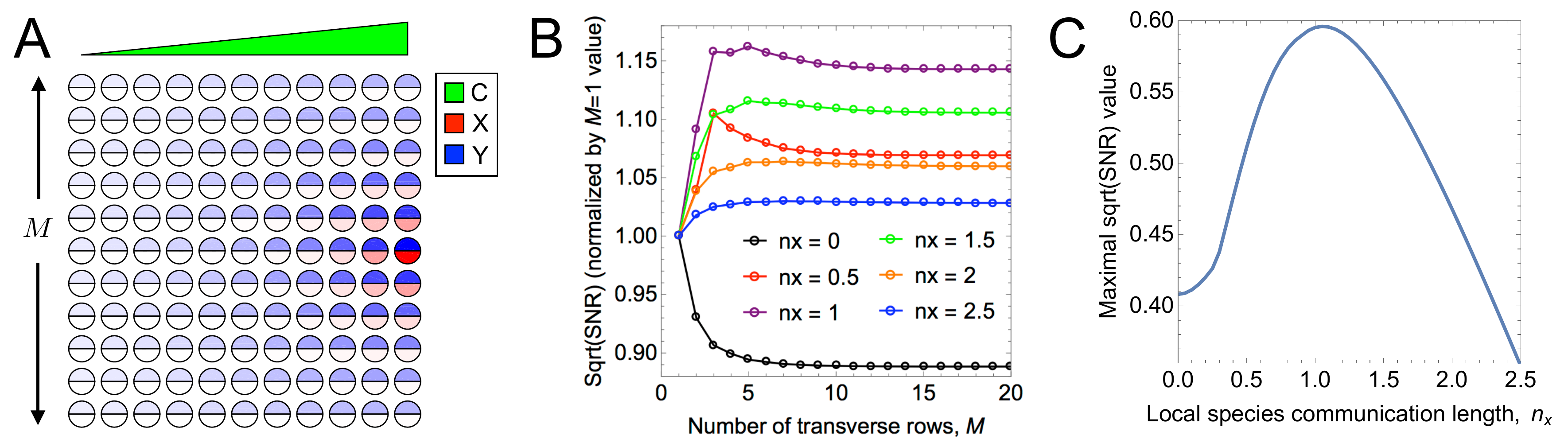}
\caption{\label{regi}
The regional excitation--global inhibition (REGI) strategy allows
cells to exploit transverse spatial averaging for gradient sensing.
(A) As in Fig.\ \ref{transverse}C, but for REGI. X molecules are
exchanged between neighboring cells, at a lower rate than Y molecules.
The difference $\Delta = x-y$ still provides the readout for the
gradient. (B) In contrast to Fig.\ \ref{transverse}D, for sufficiently
large communication length $n_x$ the SNR {\it increases} with the
number of transverse rows $M$, before ultimately decreasing, which
leads to an optimum as a function of $M$. (C) Since $n_x=0$ (LEGI) and
$n_x=n_y$ (no sensing) are suboptimal, a global optimum emerges over
both $M$ and $n_x$. Parameters are as in Fig.\ \ref{transverse}, with
$n_x = 1$ in A, which is near its optimal value according to C .}
\end{figure}

To answer this question, we once again consider a linear signal
profile, and we compute the SNR of the gradient readout $\Delta_N$
under the REGI model (see Background), as a function of the number $M$ of
rows of cells in a direction transverse to the gradient (Fig.\
\ref{regi}A). We see in Fig.\ \ref{regi}B that for a sufficiently
large value of $n_x \equiv \sqrt{\gamma_x/\mu}$, which sets the
lengthscale of spatial averaging for the local species, the precision
of gradient sensing increases with $M$. This is in contrast to the
case of LEGI, for which the precision decreases with $M$ (Fig.\
\ref{transverse}D and black curve in Fig.\ \ref{regi}B). Therefore,
the recovery of covariance between X and Y in the REGI mechanism
avoids the loss of gradient sensing precision and restores the benefit
of transverse averaging.

We also see in Fig.\ \ref{regi}B that a maximal precision emerges in
the REGI model as a function of $M$ at a particular number of rows
$M^*$. This maximum is due to the fact that the exchange of $X$, which
causes an increase in precision with $M$, and the exchange of $Y$, which causes
a decrease in precision with $M$, occur on different length scales,
$n_x < n_y$. Indeed, we see that as $n_x$ increases, the location of
the maximum $M^*$ increases concomitantly. Additionally, we see in
Fig.\ \ref{regi}C that the maximal precision value first increases
with $n_x$, then decreases with $n_x$, leading to an optimal value
$n_x^*$. This is due to the previously understood tradeoff that is
introduced when $n_x$ increases: on the one hand the variance of X is
reduced, which increases precision; on the other hand, the means of X
and Y are more similar, which decreases the precision
\cite{mugler2015limits}. Here this tradeoff is modified by the
additional benefit of increasing $n_x$, namely that it increases the
covariance of X and Y in the transverse direction, and thus further
reduces the noise in gradient sensing.

\subsection{Emergence of optimal detector shapes in two and three dimensions}

The emergence of an optimal number of transverse rows of cells, seen
in the previous section, raises the more general question of whether
there is an optimal detector shape for spatially extended gradient
sensing. This question has relevance for both 2-D and 3-D
multicellular geometries involved in gradient sensing. Is the optimal
detector shape more ``hairlike'', to maximize its extent in the
gradient direction, or more ``globular'', to exploit potential
benefits of extending along the transverse direction?

\begin{figure} \centering
  \includegraphics[width=.9\textwidth]{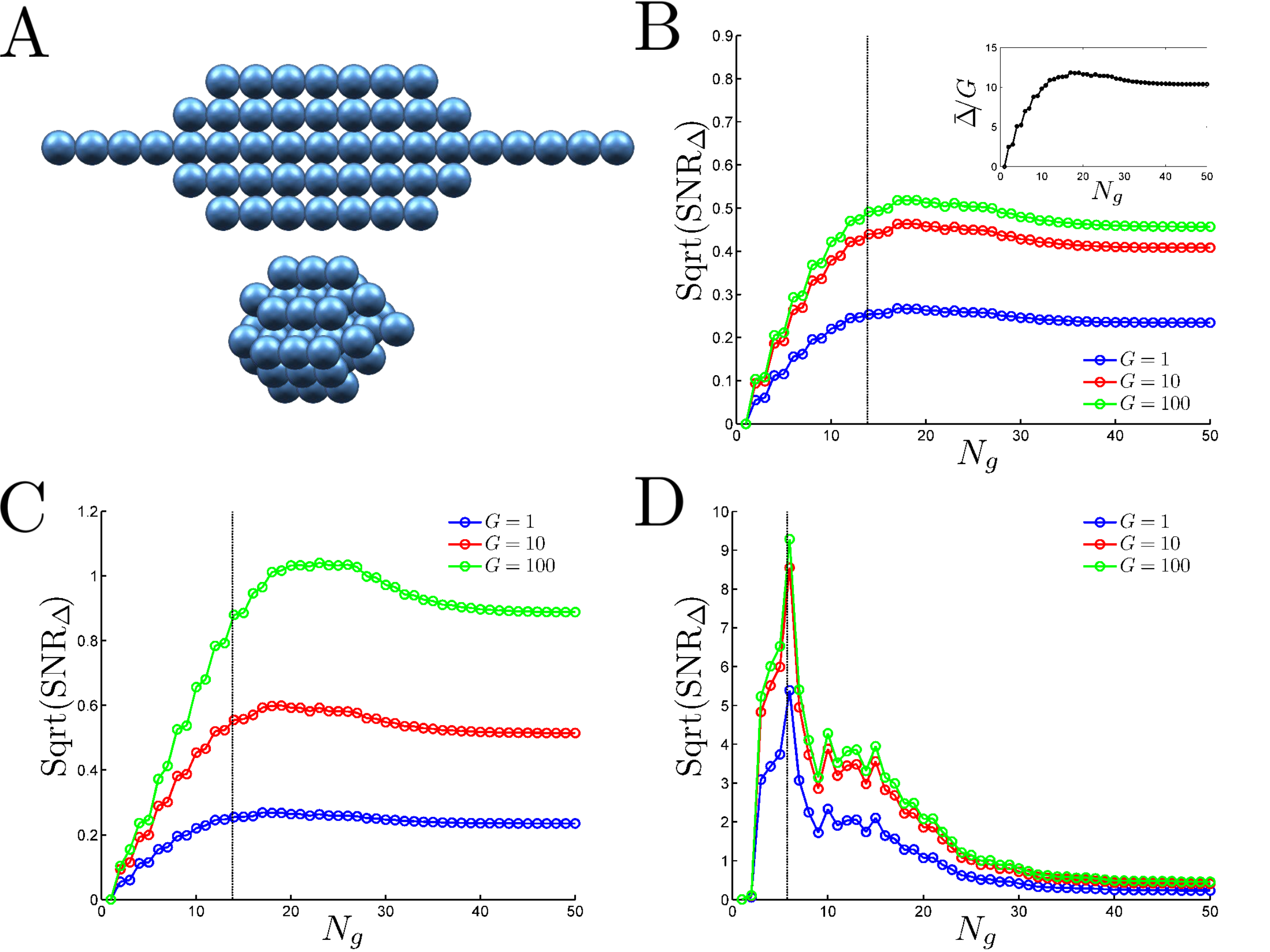} \caption{\label{opt}
    Optimal gradient sensing by 2-D and 3-D detectors. (A) Optimal
    elliptical (2-D, top) or ellipsoidal (3-D, bottom) configurations
    of $N=50$ cells for the REGI model. Gradient sensing precision is
    optimized at the rightmost cell, and the signal profile increases
    linearly to the right. We see that the optimal shapes are
    ``globular'', not ``hairlike'', especially in 3-D. (B) Precision
    vs.\ $N_g$ (the projected number of cells in the gradient
    direction) for the LEGI model in 2-D, for various gains $G$.
    Inset: mean readout $\bar{\Delta}$ normalized by $G$ (all three
    curves overlap and are colored black). (C) As in B, but for REGI.
    The additional REGI parameter $n_x$ is optimized over at each
    $N_g$ value, and the optimal precision is shown. At the observed optima in C, these values are
    $n_x^*/n_y = 0.09$ ($G=1$), $0.30$ ($G=10$), and $0.53$ ($G=100$).
    (D) As in B but for 3-D. Internal cells are shielded and do not
    sense, but do communicate. Ellipsoid axes transverse to gradient
    are equal. Optimal $n_x^* = 0$ for all $N_g$. Curve jaggedness
    arises due to numerical effects of fitting a cubic lattice of
    cells in a smooth ellipsoidal envelope. Black vertical dashed
    lines correspond to a perfect circle (B, C) or sphere (D).
    Parameters are as in Fig.\ \ref{transverse}.}
\end{figure}

To address this question, we perform a controlled optimization for
both 2-D and 3-D multicellular geometries. For a fixed number of cells
$N = 50$, we confine cells to an elliptical (2-D) or ellipsoidal (3-D)
envelope, and compute the precision of gradient sensing as a function
of the ellipse axis parameters (LEGI), as well as the ratio of
averaging length scales $n_x/n_y$ (REGI), exhaustively exploring
substantial ranges of both. In addition to the extra shape parameter,
there is one more important difference between the 2-D and 3-D cases:
in the 2-D case, we assume that every cell detects signal molecules,
since we imagine that these molecules diffuse in the 3-D
bulk, while the cells form a sensory sheet exposed to the bulk. In
contrast, in the 3-D case, we assume that only the surface cells
detect signal molecules, whereas cells that are blocked on all six
sides by neighboring cells are ``shielded'' and thus do not detect
signal molecules (although all cells still communicate via molecule
exchange). The optimal detector shapes determined by such exhaustive
search for the REGI model are shown in Fig.\ \ref{opt}A, for 2-D (top)
and 3-D (bottom).

To explain why these optimal shapes emerge, we present the precision
of gradient sensing as a function of the control parameters. First we
investigate the behavior of the LEGI model in 2-D (Fig.\ \ref{opt}B).
The control parameter is $N_g$, the (projected) number of cells in the
gradient direction, which is set uniquely in 2-D by the ratio of the
ellipse axis parameters. Small $N_g \to 1$ corresponds to a chain of
cells transverse to the gradient, while large $N_g \to N$ corresponds
to a chain of cells parallel to the gradient. The small ``stair
steps'' in the curves are due to the numerical task of fitting the
discrete multicellular square lattice within the continuous elliptical
envelope. We see that the precision vanishes at $N_g = 1$, as
expected, since in our model a single cell cannot perform gradient
detection. The precision is near maximal at $N_g = N$. This trend is
analogous to that seen for LEGI in Fig.\ \ref{transverse}D, where here
$N/N_g \sim M$ is the analog of the number of transverse rows.
However, unlike in Fig.\ \ref{transverse}D, we see in Fig.\ \ref{opt}B
that there is a weak optimum at an intermediate value of $N_g$. This
is due to a difference between the protocols of adding rows of cells
(Fig.\ \ref{transverse}D) and reshaping a fixed number of cells (Fig.\
\ref{opt}B). Adding rows does not change $\bar{\Delta}_N$. In
contrast, as seen in the inset of Fig.\ \ref{opt}B, reshaping changes
$\bar{\Delta}_N$. The reason is that elliptical configurations (like
Fig.\ \ref{opt}A, top) are not translationally symmetric in the
transverse direction. In particular, a large density of cells in the
middle of the configuration is a sink for molecules of Y. This decreases
the mean number of Y in the rightmost cell, $\bar{y}_N$, which weakly
increases the signal $\bar{\Delta}_N = \bar{x}_N - \bar{y}_N$ at
intermediate values of $N_g$ (Fig.\ \ref{opt}B inset), and therefore
increases the precision (Fig.\ \ref{opt}B). Finally, we see that the
precision increases with the gain $G$, as expected, and that the
increase saturates with $G$, since then the variance of X and Y is
dominated entirely by extrinsic, and not intrinsic, noise (see
Background).

Next we investigate the behavior of REGI in 2-D (Fig.\ \ref{opt}C).
Once again the control parameter is $N_g$. Additionally, at every
$N_g$ we optimize the local species' averaging length scale $n_x$
(generally we find an optimal value between $\sim$$0.1n_y$ and $\sim$$0.5n_y$, see
Fig.\ \ref{opt}). We see in Fig.\ \ref{opt}C that the trend of
precision versus $N_g$ is similar to that of the LEGI model (Fig.\
\ref{opt}C), but with two key differences. First, the precision is
higher for REGI than for LEGI. This is due to regional averaging
reducing the variance of the local species, as was known previously
for the 1-D model \cite{mugler2015limits}. Second, the optimum in the
precision as a function of $N_g$ is more pronounced for REGI than for
LEGI. This is because the region surrounding the optimum corresponds
to near-circular ellipses, where considerable transverse averaging
occurs. As shown in the previous section, transverse averaging
increases precision in the REGI model. Overall, the optimal structure
(Fig.\ \ref{opt}A, top) is closer to a ``globular'' circle than to
``hairlike'' chain (compare locations of the optima to the dashed vertical line in
Fig.\ \ref{opt}C, which corresponds to a perfect circle). Therefore,
we see that optimal gradient sensing by a 2-D structure benefits from
an elliptical shape in which transverse averaging occurs.

Finally, we investigate the behavior of REGI in 3-D (Fig.\
\ref{opt}D). Here there are two control parameters: the number of
cells in the gradient direction $N_g$, and the asymmetry of the
ellipsoid in the two directions transverse to the gradient. Generally
we find that the optimal shape at a fixed $N_g$ displays symmetry in
the two transverse directions, and therefore we impose this symmetry
explicitly and focus on the control parameter $N_g$. As before, at
every $N_g$ we optimize the local species' averaging length scale
$n_x$. Importantly, in the 3-D geometry, we find that the optimal
value at every $N_g$ is $n_x^* = 0$, corresponding to no averaging of
the local species (an effective LEGI model). This is due to the
shielding of internal cells: since internal cells do not detect signal
molecules, averaging of the local species would dramatically reduce
the mean local readout, making it far less than the actual local
signal value at the edge cell. This would severely reduce the mean
$\bar{\Delta}_N$, and thus the precision. The dependence of precision
on $N_g$ is shown in Fig.\ \ref{opt}D. The additional jaggedness is
again due to the incommensurate nature of the cubic cell lattice with
the smooth ellipsoidal envelope, here amplified due to the additional
dimension. We see in Fig.\ \ref{opt}D that there is again an optimum.
In fact, it is much more pronounced than in 2-D: the overall value of
the precision is ten-fold higher than in 2-D. This is again due to the
shielding of internal cells: the global species Y is averaged among
internal cells that do not produce it, which sharply decreases
$\bar{y}_N$, and thereby increases $\bar{\Delta}_N$ and thus the
precision.\footnote{Note that this particular effect of shielding
  will result in the value of $\bar{\Delta}_N$ being positive in every
  edge cell, instead of only the edge cells at the high end of the
  gradient. The sensory outcomes are still biased, but are less
  adaptive, similar to ``tug-of-war'' chemotaxis mechanisms that have
  been proposed \cite{camley2015emergent}. } Overall, the optimal
structure is very ``globular'' (Fig.\ \ref{opt}A, bottom). Indeed, it
is almost a sphere (compare the optima to the dashed vertical line in
Fig.\ \ref{opt}D). We conclude that, due to the combined effects of
spatial averaging and shielding, the optimal 3-D detector of linear
gradients extends significantly in all three spatial dimensions.

\section{Discussion}

We have investigated theoretically and computationally the ways in
which the precision of spatially extended, multi-component gradient
sensing is affected by detector geometry. Using a minimal model of
adaptive gradient sensing (LEGI), we have found that, unlike for
concentration sensing, the precision of gradient sensing decreases
with the size of the detector in a direction transverse to the
gradient. This is due to the competing effects of noise reduction and
a reduction of the covariance between concentrations subtracted to
estimate the gradient. We have demonstrated that a simple modification
of LEGI (REGI) restores the covariance and recovers the benefit of
transverse averaging for gradient sensing. The result is that the
optimal detectors in 2-D and 3-D are more globular than hairlike.

Our study elucidates the important roles of spatial averaging in
gradient sensing, which are several-fold. First, there is spatial
averaging along the gradient. In both LEGI and REGI, the global
species Y is averaged along the gradient. For a linear signal profile,
this averaging both increases the signal $\bar{\Delta}^2$, and
decreases the noise $(\delta{\Delta})^2$. Therefore, it is optimal for
Y to be averaged along the gradient to as large an extent as
possible. Second, in the REGI model, the local species X is also
averaged along the gradient. This decreases the signal but also
decreases the noise \cite{mugler2015limits}. Therefore, there is often
an optimal ratio $n_x/n_y$ of the spatial extents of the
averaging. Third, there is spatial averaging transverse to the
gradient. In the LEGI model, only Y is averaged transverse to the
gradient. In a translationally symmetric geometry, this does not
change the signal, but it changes the noise by both decreasing the
variance of Y and decreasing the covariance between X and Y. These
have opposite effects on the precision. For LEGI, the latter
dominates, decreasing the precision. Therefore, transverse averaging
is detrimental for gradient sensing. However, in the REGI model, X is
also averaged transverse to the gradient. Once again, in a
translationally symmetric geometry, this does not change the signal
with respect to REGI in 1-D, but it decreases the noise, both by
further reducing the variance in X and by restoring a larger
covariance between X and Y. Therefore, transverse averaging is
beneficial for REGI-type gradient sensing. These roles of spatial
averaging are modified in geometries without translational symmetry as
we discussed above.  However, the net result remains the same: the
optimal 2-D and 3-D REGI-type gradient detectors are globular,
benefitting from extensive spatial averaging in the transverse
directions.

Our study also reveals the effects of shielding of signal from the
inner cells in a 3-D geometry. Shielding amplifies the effect of
spatial averaging, since the measurements performed by edge cells,
which detect signal, are averaged with those of their interior
neighbors, which do not detect signal. This amplification increases
the signal in a particular edge cell, but makes the system less
adaptive, since every edge cell has an above-average readout. With
shielding, a more appropriate measure of the sensory outcome might
therefore be the difference in readouts between cells up and down the
gradient, e.~g.\ $\bar{\Delta}_N-\bar{\Delta}_1$. This measure is
likely to depend nontrivially on internal and geometric parameters
such as $n_x$ and $M$, and will likely result in a nontrivial optimal
local averaging length scale, $n_x^*\neq 0$. Another possibility is
that gain $G$ should be different in Eqs.~\ref{xgain} and \ref{ygain},
compensating for the two messenger molecules averaging over different
numbers of neighbors that do not detect the ligand.  We leave both of
these interesting explorations for future investigations.

In this work, we have emphasized the distinction between (i)
concentration sensing within a graded concentration profile and (ii)
gradient sensing. For example, in Drosophila development, individual
nuclei in the embryo measure (and are thought to threshold) the local
concentration, even though the morphogen gradient is graded
\cite{Gregor:2007du,erdmann2009role,Lander:2011bj,Sokolowski:2015tn}. This
is an example of concentration sensing.  In contrast, gradient
sensing, as explored here, is the task of obtaining an internal
readout of the {\it difference} in local signal concentrations at two
or more different points in space. In other words, unlike
concentration sensing, gradient sensing determines the direction in
which the concentration changes, and it allows subsequent directional
polarization of the sensor.  This definition of gradient sensing, by
construction, is adaptive: the readout does not depend on the
background concentration. Systems that respond adaptively and
directionally to chemical gradients, such as amoeba \cite{amoeba} and
epithelial cell groups \cite{ellison2015cell}, are preforming gradient
sensing. Because concentration sensing and gradient sensing are
distinct, it may not be so surprising that transverse averaging has
very different effects on them: the precision of concentration sensing
increases with the transverse size, whereas the precision of LEGI
gradient sensing decreases with the transverse size (Fig.\
\ref{transverse}).

How do our results compare to experimental systems? A well-studied
example of a natural gradient-sensing system is the growth
factor-directed extension of mammary epithelial ducts
\cite{ewald2008collective,huebner2014cellular}. Gradient sensing in
this system has been shown to be multicellular and adaptive
\cite{ellison2015cell}. {\it In vivo}, the extension is led by an
``end bud'' of cells at the duct tip. These tips can form either long
hairlike structures or coalesce into nearly spherical globules, as was
observed in organotypic studies with different chemical and genetic
perturbations \cite{ellison2015cell}. Long hairs could act as
``feelers'' for the duct, sampling a long swath of the environment in
the gradient direction. However, our analysis predicts that such
hairlike morphologies are suboptimal, and the globular bud shape, as
in Fig.\ \ref{opt}A, would produce a better precision. In agreement
with the prediction, the end buds in wildtype mice are nearly
spherical, and the globule is often wider than the duct itself
\cite{huebner2014cellular}. Similarly, neither chemotaxing amoeba
\cite{amoeba} and neutrophils \cite{Onsum:2006gg}, nor growing neurons
\cite{Rosoff2004} form very thin hairlike protrusions to facilitate
sensing. Instead they keep the aspect ratio of the gradient sensing
part of the protrusions closer to one, again supporting our
findings. Further, in Drosophila border cell migration, another
example of directional collective cell behavior, groups of cells
travel as a sphere in a confined space, where it would have been
easier to travel as a chain \cite{Bianco:2007hi}. All of these
examples provide indirect evidence that transverse averaging is used
in multiple biological contexts. While direct tests of effects of
transverse averaging have not been done, they are certainly
possible. Indeed, as mentioned above, different perturbations to
organotypic epithelial cultures result in them assuming different
geometric shapes \cite{ellison2015cell}. Thus it should be possible to
measure the accuracy of sensing (and the subsequent organoid
polarization) as a function of the shape. Such experiments would allow
direct testing of our main prediction that transverse averaging leads
to more accurate directional sensing outcomes, especially in REGI-type
models.

\section*{Acknowledgements}
TS was supported by the Woodruff Fellowship at Emory University. SF
and AM were supported in part by Simons Foundation grant 376198. AL
was supported in part by NSF grant PoLS-1410545. IN was supported in
part by NSF grant PoLS-1410978.

\section*{References}

\providecommand{\newblock}{}

\end{document}